\documentclass[aps,prl,twocolumn,nofootinbib,a4paper,floatfix,10pt]{revtex4-1}

\usepackage{color}
\usepackage{bm}
\usepackage{amssymb,amsmath,color,graphicx}
\usepackage{amsfonts}
\usepackage{dsfont}
\usepackage{bbold}
\usepackage{graphicx}
\usepackage{float}
\usepackage{subfigure}
\usepackage{verbatim}
\usepackage{amsfonts}
\usepackage{dcolumn}
\usepackage{bm}
\usepackage{color}
\usepackage[dvipsnames]{xcolor}
\usepackage{listings}
\definecolor{zaffre}{rgb}{0.0, 0.08, 0.66}
\usepackage[colorlinks=true,urlcolor=zaffre,citecolor=zaffre,linkcolor=zaffre]{hyperref}
\usepackage{mathrsfs}
\usepackage{filecontents}

\makeatletter
\def\maketag@@@#1{\hbox{\m@th\normalfont\normalsize#1}}  
\makeatother

\begin{document}

\title{Optimal realistic attacks in continuous-variable quantum key distribution}
\author{Nedasadat Hosseinidehaj$^1$}  \email{n.hosseinidehaj@uq.edu.au}

\author{Nathan Walk$^{2,3}$}  \email{nwalk@zedat.fu-berlin.de}

\author{Timothy C. Ralph$^1$}  \email{ralph@physics.uq.edu.au}

\affiliation{$^1$Centre for Quantum Computation and Communication Technology, School of Mathematics and Physics, University of Queensland, St Lucia, Queensland 4072, Australia.}

\affiliation{$^2$Dahlem Center for Complex Quantum Systems, Freie Universit{\"a}t Berlin, 14195 Berlin, Germany.}
\affiliation{$^3$Department of Computer Science, University of Oxford, Wolfson Building, Parks Road, Oxford OX1 3QD, United Kingdom.}


\date{\today}

\begin{abstract}

Quantum cryptographic protocols are typically analysed by assuming that potential opponents can carry out all physical operations, an assumption which grants capabilities far in excess of present technology. Adjusting this assumption to reflect more realistic capabilities is an attractive prospect, but one that can only be justified with a rigorous, quantitative framework that relates adversarial restrictions to the protocols security and performance. We investigate the effect of limitations on the eavesdropper's (Eve's) ability to make a coherent attack on the security of continuous-variable quantum key distribution (CV-QKD). We consider a realistic attack, in which the total decoherence induced during the attack is modelled by a Gaussian channel. Based on our decoherence model we propose an optimal hybrid attack, which allows Eve to perform a combination of both coherent and individual attacks simultaneously. We evaluate the asymptotic and composable finite-size security of a heterodyne CV-QKD protocol against such hybrid attacks in terms of Eve's decoherence. We show that when the decoherence is greater than a threshold value, Eve's most effective strategy is reduced to the individual attack. Thus, if we are willing to assume that the decoherence caused by the memory and the collective measurement is large enough, it is sufficient to analyse the security of the protocol only against individual attacks, which significantly improves the CV-QKD performance in terms of both the key rate and the maximum secure transmission distance.
\end{abstract}

\maketitle

Coherent attacks are known to be the most powerful evesdropping attacks on quantum key distribution (QKD) protocols.  The evesdropper, Eve, prepares a global ancillary system, interacting collectively with all the quantum states sent through the channel, with the entire output ancillae stored into a quantum memory, and a collective measurement applied over the stored ensemble \cite{thesis, Weedbrook2012} to extract the maximum information on the key. Making such an attack, particularly on a continuous-variable (CV) QKD system \cite{thesis, Weedbrook2012, Diamanti} represents an extreme technical challenge for Eve.

For a no-switching CV-QKD protocol \cite{no-switching1,no-switching2}, based on Gaussian-modulated coherent states and heterodyne detection, the finite-size composable security against coherent attacks can be analysed by considering Gaussian collective attacks \cite{Finite-size-Leverrier2017}. In a collective attack Eve prepares an ensemble of independent and identical quantum systems, each one interacting individually with a single quantum state transmitted through the channel, with the output ancilla stored into a quantum memory \cite{thesis, Weedbrook2012}. In a collective attack on Gaussian-modulated coherent-state CV-QKD protocol the ancillae stored in Eve's quantum memory is a tensor product of $n$ coherent states, i.e., an $n$-symbol codeword. In order for Eve to extract the maximum information upper bounded by the Holevo information \cite{Holevo1,Holevo2,Holevo3} a sequence of projective binary-outcome collective quantum measurements has to be applied to the $n$-symbol codeword \cite{PRA-sequential2012}. In \cite{IEEE-sequential2012} a quantum optical realization of the sequential decoding strategy has been provided, which in a large number of $2^{nR}$ steps determines which codeword was sent (with $R$ the rate in bits/symbol being bounded by the the Holevo information). In \cite{sequential2016} a more efficient (in terms of scaling) sequential decoding strategy (but with no evidence of quantum optical implementation) has been proposed, consisting of a sequence of complex adaptive collective quantum measurements performed in a series of $nR$ concatenated steps to determine which codeword was sent. Thus, in a realistic collective attack a significant amount of time and/or coherent operations are required for Eve to collectively decode the stored ensemble to approach the Holevo information.


In this work we investigate Eve's optimal attack in a no-switching CV-QKD protocol, given practical restrictions on her storage and processing ability. The realistic assumption of restricted quantum memories has been studied in the context of quantum data-locking protocols \cite{QDL2014, QDL2015-njp, QDL2015-pra, QDL2015-review}, and two-party cryptographic tasks of oblivious transfer and bit commitment \cite{SW2008, SW2010, IEEE2012, exp-DV-2014,Furrer-noisy-QM}. In a no-switching CV-QKD protocol Eve can avoid the decoherence induced over the storage and processing time of the collective attack by performing individual attacks, where she interacts individually with each quantum state sent by Alice, and she immediately performs an individual measurement on the output ancilla. This is an optimal individual attack strategy because there is no basis information withheld in the no-switching protocol. With the aim of allowing Eve to simultaneously benefit from both the collective decoding and avoiding the decoherence induced over the decoding, we will propose a new class of optimal attacks, hybrid attacks, that lie in between the coherent and individual attacks. In the hybrid attack we model the total decoherence induced on each quantum system stored into the quantum memory with a thermal, lossy Gaussian channel. We will evaluate the asymptotic and composable finite-size security of a no-switching CV-QKD protocol in terms of Eve's attack decoherence, thereby demonstrating that if the decoherence is higher than a threshold value, Eve's best strategy is the individual attack, and thus the security of the CV-QKD protocol can be analysed by considering only the individual attack, which remarkably improves both the key rate and the maximum secure transmission distance of the protocol. Note that our realistic assumption of decoherence over the storage time of a collective (or coherent) attack is fully future proof, in the sense that if a perfect quantum memory becomes possible in the future, the key which is secure now will remain secure.

\paragraph*{CV-QKD system.} We consider a Gaussian no-switching CV-QKD protocol \cite{no-switching1,no-switching2}, where Alice generates a pair of random real numbers, chosen from two independent Gaussian distributions of variance $V_A$, to prepare coherent states. The prepared states are then transmitted over an insecure quantum channel with transmissivity $T$ and excess noise $\xi$ (relative to the input of the quantum channel) to Bob. For each incoming state, Bob uses heterodyne detection to measure both the $\hat q$ and $\hat p$ quadratures. In this protocol, sifting is not needed, since both of the random variables generated by Alice are used for the generation of the key. When the quantum communication is finished and all the incoming quantum states are measured by Bob, classical post-processing including discretization, parameter estimation, error correction, and privacy amplification over a public but authenticated classical channel is commenced to produce a shared secret key.

This Gaussian CV-QKD system can also be represented by an equivalent entanglement-based scheme \cite{thesis, Weedbrook2012}, where Alice generates a two-mode squeezed vacuum (TMSV) state $\rho_{AB}$ with the quadrature variance $V{=}V_A{+}1$. Alice retains mode~$A$, while sending mode~$B$ to Bob over the quantum channel. In the entanglement-based scheme, Alice applies a heterodyne detection to mode~$A$, which results in projecting the mode~$B$ onto a coherent state. At the output of the channel, Bob applies a heterodyne detection to the received mode~$B_1$, with his detector having an efficiency of $\eta$ and electronic noise variance of $\upsilon_{\rm el}$ \cite{inefficient_homodyne, inefficient_heterodyne}.

\paragraph*{Composable finite-size security analysis.} We exploit the approach introduced in \cite{Finite-size-Leverrier,Finite-size-Leverrier2017} to analyse the composable finite-size security of the no-switching CV-QKD protocol (acting on $2n$-mode state shared between Alice and Bob) against coherent attacks. This approach consists of two steps; first proving the security of the protocol against Gaussian collective attacks with a security parameter $\epsilon$ \cite{Finite-size-Leverrier}, and then applying the Gaussian de Finetti reduction \cite{Finite-size-Leverrier2017} to obtain the security against coherent attacks with a polynomially larger security parameter $\tilde{\epsilon}$ \cite{Finite-size-Leverrier2017}, where the security loss due to the reduction from coherent attacks to collective attacks scales like $\mathit{O}(n^4)$ (see Appendix.~A for more details). There exists another approach to prove the security against coherent attacks which is based on an entropic uncertainty relation \cite{Finite-size-Furrer, Finite-size-Furrer-RR, Finite-size-exp}, but the relevant CV-QKD protocol requires the preparation of squeezed states. Furthermore, due to the looseness of the current best entropic uncertainty relations, this approach predicts key rates that are pessimistic as a function of loss.

The no-switching CV-QKD protocol acting on $2n$-mode state shared between Alice and Bob is $\epsilon$-secure against collective attacks in a reverse reconciliation (RR) scenario if $\epsilon {=} 2\epsilon_{\rm sm}{+}\bar \epsilon {+}\epsilon_{\rm PE}{+}\epsilon_{\rm cor}$ \cite{Finite-size-Leverrier,Finite-size-Lupo} and if the key length $\ell$ is chosen such that \cite{Finite-size-Leverrier,Finite-size-Lupo}
\begin{equation}\label{key-length-main}
\begin{array}{l}
 \ell  {\le} N[ \beta I(A{:}B) {-} \chi(B{:}E) ] {-} {\Delta _{\rm AEP}} {-} 2\log_2 (\frac{1}{{2\bar \epsilon }}),
 \end{array}
\end{equation}
where $I(A{:}B)$ is the Shannon mutual information between Alice and Bob (calculated and provided in Appendix.~B), $\chi(B{:}E)$ is the Holevo mutual information between Eve and Bob, $\beta$ is the reconciliation efficiency, $N{=}2n$, $\Delta_{\rm AEP} = \sqrt N [(d{+}1)^2{+}4(d{+}1)\sqrt{\log_2({2{/}\epsilon_{\rm sm}^2})} {+} 2\log_2({2}{/}({\epsilon^2 \epsilon_{\rm sm}})) ] {+} 4{\epsilon_{\rm sm}d}{/}{\epsilon}$ \cite{Finite-size-Leverrier,Finite-size-Lupo}, and where $d$ is the discretization parameter, and $\epsilon_{\rm cor}$ and $\epsilon_{\rm PE}$ are the maximum failure probabilities for the error correction and parameter estimation, respectively (see Appendix.~A for more details). We have considered the same scenario as \cite{Finite-size-Leverrier,Finite-size-Lupo, PE-MDI-2018}, where almost all the raw data can be utilized to distill the secret key. Note that for the $\epsilon$-security analysis of the same protocol against individual attacks we can still use Eq.~(\ref{key-length-main}), where $\chi(B{:}E)$ must be replaced by the Shannon mutual information between Eve and Bob, $I(B{:}E)$.

\begin{figure}[b!]
    \begin{center}
      {\includegraphics[width=3.2 in, height=2 in]{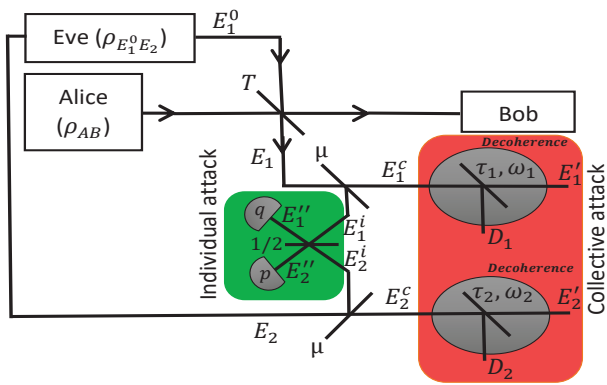}}
    \caption{Eve's optimal hybrid attack for the no-switching CV-QKD protocol.}\label{QM}
    \end{center}
\end{figure}

\paragraph*{Optimal realistic attack.} Now we investigate the optimal eavesdropping attack on a no-switching CV-QKD protocol, given Eve's storage and processing limitations. We propose an optimal realistic hybrid attack, where Eve performs a combination of both the coherent and individual attacks. Note that Eve does not need a quantum memory for the individual attack, since she does not need to wait for any basis information to be disclosed in the no-switching CV-QKD protocol. This hybrid attack allows Eve to benefit from the advantage of both the collective decoding, as well as the individual measurement of the non-decohered ancillae.
We model the coherent-attack (individual-attack) part of the hybrid attack with a Gaussian collective (individual) attack. Gaussian collective (individual) attacks are known to be asymptotically optimal \cite{Gaussian-optimality-1, Gaussian-optimality-2, Gaussian-optimality-3, Renner2009, thesis, Weedbrook2012}. Furthermore, according to the Gaussian de Finetti reduction, for the no-switching protocol it is also sufficient to consider Gaussian collective attacks in the finite-size, composable security proof \cite{Finite-size-Leverrier2017}. These results are crucial since they allow us to explicitly model Eve's attack and her decoherence. Both the optimal Gaussian collective attack \cite{collective} and the optimal Gaussian individual attack \cite{individula-2007-1} can be modelled using an entangling cloner attack (shown in Fig.~\ref{QM}), where Eve replaces the Gaussian channel with transmissivity $T$ and excess noise $\xi$ by a TMSV state $\rho_{E^0_1E2}$ of the quadrature variance $\omega_E {=} 1 {+} T \xi{/}(1{-}T)$, and a beam splitter of transmissivity $T$.
Half of the TMSV state, mode $E^0_1$, is mixed with the state sent by Alice in the beam splitter, outputting mode $B_1$ (which is sent to Bob through a perfect quantum channel) and Eve's ancillary, mode $E_1$.

In order to combine both the Gaussian collective attack and the Gaussian individual attack in a hybrid attack, we exploit two beam splitters with identical transmissivities $\mu$ to split each of Eve's ancillary modes into two output modes, one for the collective attack and the other one for the individual attack. In fact, the output mode $E_1$ ($E_2$) is split in a beam splitter of transmissivity $\mu$ into two output modes $E^c_1$ ($E^c_2$) for the collective attack and $E^i_1$ ($E^i_2$) for the individual attack. The ancillary modes $E^c_1$ and $E^c_2$ are stored into Eve's quantum memories, and collectively measured after the entire ancillae are stored. 
Since we are modelling Gaussian attacks, we model the total decoherence induced during the collective attack over the storage and processing time by a thermal, lossy Gaussian channel with transmissivity $\tau$ and thermal noise variance $\omega$.
Explicitly, the ancillary mode $E^c_1$ ($E^c_2$) undergoes the decoherence, modelled by a Gaussian channel with parameters $\tau_1, \omega_1$ ($\tau_2, \omega_2$), and the output mode $E'_1$ ($E'_2$). Note that the output modes $D_1$  and $D_2$ are not accessible to Eve.
On the other hand, in the individual attack, the ancillary modes $E^i_1$ and $E^i_2$ are mixed in a balanced beam splitter resulting in modes $E''_1$ and $E''_2$, where the $\hat q$ quadrature (the $\hat p$ quadrature) is measured using the homodyne detection on $E''_1$ ($E''_2$) \cite{individula-2007-1,individula-2007-2}.

\paragraph*{Security analysis against the hybrid attack.} The finite-size key length of the no switching protocol in the RR scenario which is secure against the hybrid attack with the security parameter $\tilde \epsilon$ can be given as
\begin{equation}\label{key-length-hybrid}
\begin{array}{l}
 \ell_{\rm hyb}  {\le}
 \mathop {\min }\limits_\mu [N \beta I(A{:}B) {-}   N I_\mu^{\rm hyb}(B{:}E)
{-} {\Delta^{H} _{\rm AEP}} {-} 2\log_2 (\frac{1}{{2\bar \epsilon }})],
 \end{array}
\end{equation}
where $I_\mu^{\rm hyb}(B{:}E)$ is the upper bound on the mutual information between Eve and Bob, which is given by 
\begin{equation}\label{mutual-hybrid}
\begin{array}{l}
I_\mu^{\rm hyb}(B{:}E) = I_\mu^{\rm hyb}(B{:}E'_1E'_2 E''_1E''_2 ) =
\\
\chi_\mu(BE''_1E''_2{:}E'_1E'_2) + I_\mu(B{:}E''_1E''_2) - \chi_\mu(E''_1E''_2{:}E'_1E'_2),
 \end{array}
\end{equation}
where $\chi_\mu(BE''_1E''_2{:}E'_1E'_2)$ is Eve's information contributed from the collective attack, limited by the Holevo bound, $I_\mu(B{:}E''_1E''_2)$ is Eve's information contributed from the individual attack, limited by the Shannon bound, and $\chi_\mu(E''_1E''_2{:}E'_1E'_2)$ is the mutual information between Eve's ancillary modes for the individual and collective attacks, limited by the Holevo bound. See Appendix.~C for the calculation of the right-hand terms of Eq.~(\ref{mutual-hybrid}).

Since $\Delta$-term in Eq.~(\ref{key-length-hybrid}) is different for the coherent and individual attacks, to compute $\ell_{\rm hyb}$ in Eq.~(\ref{key-length-hybrid}) we first maximise $I_\mu^{\rm hyb}(B{:}E)$ over all possible values of $0 {\le} \mu {\le} 1$. When the maximization of $I_\mu^{\rm hyb}(B{:}E)$ leads to $\mu{=}1$, Eve's hybrid attack reduces to the coherent attack.
In this case, Eq.~(\ref{key-length-hybrid}) changes to $\ell_{\rm hyb} {=} \ell_{\rm coh}$, where
\begin{equation}\label{key-length-coherent}
\begin{array}{l}
 \ell_{\rm coh}  {\le} N\left[ \beta I(A{:}B) {-} \chi(B{:}E_1'E_2') \right] {-} {\Delta^{C} _{\rm AEP}} {-} 2\log_2 (\frac{1}{{2\bar \epsilon }}),
 \end{array}
\end{equation}
and where $\chi(B{:}E_1'E_2'){=}\chi_\mu(BE''_1E''_2{:}E'_1E'_2)$ for $\mu{=}1$, and ${\Delta^{C} _{\rm AEP}}$ is given by $\Delta_{\rm AEP}$ in Eq.~(\ref{key-length-main}) for $\epsilon {\ll} \tilde \epsilon $.
When the maximization of $I_\mu^{\rm hyb}(B{:}E)$ leads to $\mu{=}0$, Eve's hybrid attack always reduces to the individual attack in the asymptotic regime. However, it is not the case for the finite-size regime, since $\Delta$-term for the coherent attack is much larger than that of the individual attack. 
This is because the $\Delta$-term for the individual attack does not have to include the $\mathit{O}(n^4)$ reduction in $\tilde \epsilon$ that is required to reduce coherent attacks to collective ones. This means there are instances where, although the coherent attack results in a smaller mutual information with Eve (which we would associate with a higher asymptotic key rate), the coherent-attack finite key rate is still lower than the individual-attack finite key rate because of this difference in the finite-size corrections. 
Hence, when the maximization of $I_\mu^{\rm hyb}(B{:}E)$ leads to $\mu{=}0$, the finite-size key length is obtained by $\ell_{\rm hyb} {=} \min (\ell_{\rm coh}{,}\ell_{\rm ind}) \rm$, where $\ell_{\rm ind}$ is the finite-size key length where Eve's hybrid attack reduces to the individual attack, and is given by
\begin{equation}\label{key-length-individual}
\begin{array}{l}
 \ell_{\rm ind}  {\le} N\left[ \beta I(A{:}B) {-} I(B{:}E''_1E''_2) \right]
 {-} {\Delta^{I} _{\rm AEP}} {-} 2\log_2 (\frac{1}{{2\bar \epsilon }}),
 \end{array}
\end{equation}
and where $I(B{:}E''_1E''_2){=}I_\mu(B{:}E''_1E''_2)$ for $\mu{=}0$, and ${\Delta^{I} _{\rm AEP}}$ is given by $\Delta_{\rm AEP}$ in Eq.~(\ref{key-length-main}) for $\epsilon {=} \tilde \epsilon$. Furthermore, when the maximization of $I_\mu^{\rm hyb}(B{:}E)$ leads to $0{<}\mu{<}1$, Eve performs a combination of both the individual and coherent attacks. In this case we can only calculate a (presumably loose) lower bound on the finite-size key length $\ell_{\rm hyb}$. Since $\epsilon {\ll} \tilde \epsilon$ leads to ${\Delta^{C} _{\rm AEP}} {>} {\Delta^{I} _{\rm AEP}}$, the (loose) lower bound on $\ell_{\rm hyb}$ can be obtained by Eq.~(\ref{key-length-hybrid}) where ${\Delta^{H} _{\rm AEP}}{=}{\Delta^{C} _{\rm AEP}}$.

\begin{figure}[h!]
    \begin{center}
      {\includegraphics[width=3.2 in, height=2.5 in]{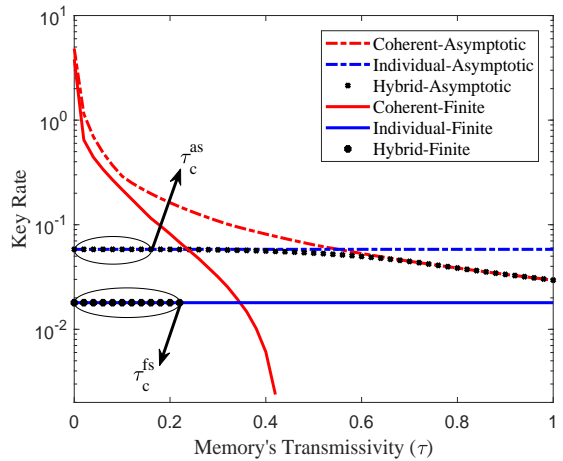}}
    \caption{The finite-size and asymptotic key rate as a function of memory's transmissivity, $\tau$, for individual (blue), coherent (red), and hybrid attacks (black). The numerical values are $\eta{=}0.6$ \cite{experiment-CVQKD-2016}, $\upsilon_{\rm el}{=}0.015$ \cite{experiment-CVQKD-2013}, $T{=}0.1$, $\xi{=}0.01$, $d{=}5$, $\omega{=}1, \beta{=}0.98$ \cite{beta-0.98},  and the modulation variance is optimized. The region marked by the ellipse shows memory's transmissivities for which Eve's optimal attack is the individual attack.}\label{FS-main}
    \end{center}
\end{figure}

\paragraph*{Numerical results.} We illustrate these results with a practical example of realistic devices \cite{experiment-CVQKD-2013,experiment-CVQKD-2016} and a lossy channel with transmissivity $T=0.1$ (or approximately 50km of telecom fibre) and $\xi=0.01$. In Fig.~\ref{FS-main} the asymptotic and finite-size key rate of the no-switching protocol in the RR scenario is illustrated as a function of Eve's memory-channel transmissivity for different types of attacks; individual, coherent, and hybrid attacks. In the asymptotic regime the secret key rate in the RR scenario is given by $K_{\rm ind} {=} \beta I(A{:}B) {-} I(B{:}E''_1E''_2)$ against the individual attack, $K_{\rm col} {=} \beta I(A{:}B) {-} \chi(B{:}E_1'E_2')$ against the collective attack, and $K_{\rm hyb} {=} \beta I(A{:}B) {-} \mathop {\max }\limits_\mu  [I_\mu^{\rm hyb}(B{:}E)] $ against the hybrid attack. Note that the derived bounds for the secret key rate in the case of collective attacks remain asymptotically valid for the arbitrary coherent attacks \cite{Renner2009}. In the finite-size regime the secret key rate in the RR scenario is given by $K_{\rm ind} {=} \frac{\ell_{\rm ind}}{N}$ against the individual attack, $K_{\rm coh} {=} \frac{\ell_{\rm coh}}{N}$ against the coherent attack, and $K_{\rm hyb} {=} \frac{\ell_{\rm hyb}}{N}$ against the hybrid attack. For all types of attacks the finite-size key is secure with the security parameter $\tilde \epsilon{=}10^{-6} $ for $n{=}10^{9}$. Recall again that to analyse the security of the no-switching protocol against coherent attacks with the security parameter $\tilde \epsilon{=}10^{-6} $, it is sufficient to analyse the security of the protocol against Gaussian collective attacks with the security parameter $\epsilon$ using Eq.~(\ref{key-length-coherent}) where $\epsilon \ll \tilde \epsilon $. Here we consider $\epsilon {=} 10^{-42}$ for $n{=}10^{9}$ and $\tilde \epsilon{=}10^{-6}$, since the security loss due to the reduction from coherent attacks to collective attacks scales like $\mathit{O}(n^4)$. In Fig.~\ref{FS-main} we assume identical (i.e. $\tau_1{=}\tau_2{=}\tau$ and $\omega_1{=}\omega_2{=}\omega$) but independent quantum memories.



In Fig.~\ref{FS-main} we see that there is a threshold transmissivity of Eve's memory channel below which an individual attack is always optimal. We denote $\tau_c^{\rm as}$ and $\tau_c^{\rm fs}$ for the threshold transmissivity in the asymptotic and finite-size regime respectively. Asymptotically we see that when $\tau{\le}\tau_c^{\rm as} {=} 0.17$ individual attacks are optimal, for $\tau_c^{\rm as}{<}\tau{\le}0.72$ Eve's optimal strategy is a hybrid combination of both individual and coherent attacks, and for $\tau{>}0.72$ coherent attacks are optimal. In the finite-size case when $\tau{\le}\tau^{\rm fs}_{c}{=}0.23$ individual attacks are optimal, and for $\tau{>}\tau_c^{\rm fs}$ (where hybrid attacks are optimal for $\tau_c^{\rm fs}{<}\tau{\le}0.7$, and coherent attacks are optimal for $\tau{>}0.7$) positive finite key rate cannot be generated. Note that in the presence of memory's thermal noise (i.e. $\omega{>}1$) the threshold transmissivity becomes higher than that in the case of Eve's pure-loss quantum memory (i.e. $\omega{=}1$). See Appendix.~D for the numerical results on low-loss channels as well as direct reconciliation scenario.


\begin{figure}[h!]
    \begin{center}
      {\includegraphics[width=3 in, height=1.5 in]{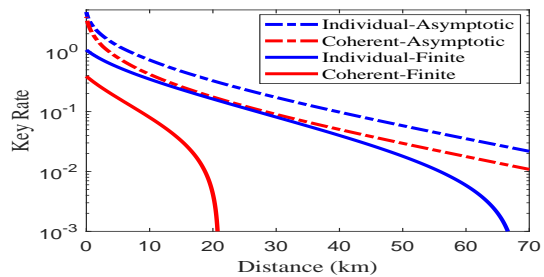}}
    \caption{The finite-size and asymptotic key rate as a function of channel distance (with the assumption of 0.2dB loss per kilometer) for individual (blue) and perfect coherent attacks (red). The numerical values are the same as Fig.~\ref{FS-main}.}\label{distance}
    \end{center}
\end{figure}

Thus, we find that our analysis can translate a model for the decoherence of Eve's attack into a rigorous, quantifiable bound on performance. This fact results in a remarkable improvement of the key rate up to that achievable under the assumption of individual attacks. For a Gaussian-channel model we generically find a threshold value for the overall decoherence of Eve's attack above which the mutual information between Bob and Eve is degraded so severely that Eve is forced to make an individual attack. These results are of significant practical relevance. 
For instance, Fig.~\ref{FS-main} shows that while positive finite key rates cannot be generated under the unrealistic assumption of perfect coherent attacks, by considering Eve's attack decoherence we are able to move from insecure regime to secure regime, and generate non-trivial positive finite key rates. Fig.~\ref{distance} also shows the advantage of individual attacks over perfect coherent attacks in terms of the maximum secure transmission distance of the CV-QKD protocol, where this advantage is significant, especially in the finite-size regime.


\paragraph*{Discussion.} In our model of restricted attack we make no assumption on the size of Eve's quantum memory. In fact, we assume a less restricted assumption on Eve's storage ability where she is able to store \emph{all} the ancillary modes. However, we assume any mode stored into the memory undergoes the same amount of decoherence. It could also be reasonable to consider a bounded memory, where only a small fraction of the total modes can be stored, and the rest of them are only individually measured. Further, it would be more realistic to consider different amounts of decoherence for Eve's stored ancillary modes, as some of them are stored in the memory longer than others. Finally, it would be desirable to extend this result to the other Gaussian CV-QKD protocols, although this would require solving the open problem of explicitly identifying the corresponding optimal attacks.

\paragraph*{Conclusion.} Given the realistic restriction that in a coherent (or collective) attack, Eve's quantum system undergoes a certain amount of decoherence over the storage and processing time, we found that there is always a threshold for Eve's decoherence, above which Eve's best strategy is limited to individual attacks. Since the decoherence is an increasing function of the storage time, if Eve's required time to store the entire ensemble and perform a collective measurement on the stored ensemble is sufficiently long, the security analysis of the protocol reduces to that of individual attacks, which substantially improves the key rate and the secure transmission distance of the CV-QKD protocol.

\paragraph*{Acknowledgements.} The authors gratefully acknowledge valuable discussions with Andrew Lance, Thomas Symul, and Helen M. Chrzanowski. This research was supported by Australian Defence Innovation Hub Funding CTD-21-2017-6. This research is also supported by the Australian Research Council (ARC) under the Centre of Excellence for Quantum Computation and Communication Technology (Project No. CE170100012). NW acknowledges funding support from the EPSRC National Quantum Technology Hub in Networked Quantum Information Technologies and funding from the European Union's Horizon 2020 research and innovation programme under the Marie Sklodowska-Curie grant agreement No. 750905.

\newpage
\appendix

\setcounter{equation}{0}
\setcounter{figure}{0}
\renewcommand{\theequation}{A\arabic{equation}}
\renewcommand{\thefigure}{A\arabic{figure}}

\section{Appendix}

\section{Appendix A: Composable finite-size security analysis}\label{APP.A}

In the finite-size security analysis the key is proved to be secure against Eve's attacks up to a small failure probability, while in the asymptotic security analysis the key is proved to be perfectly secure in the limit of infinite quantum states distributed between Alice and Bob.

The security of a quantum key distribution (QKD) protocol against general attacks is established by proving that the \emph{real} protocol is approximately equal to an \emph{ideal} protocol. We first introduce the properties an ideal protocol is required to achieve, correctness, secrecy, and robustness. Note that an entanglement-based QKD protocol can be described as a completely positive trace-preserving map that takes an input state $\rho_{AB}$, and outputs a key consisting of two classical strings $S_A$ and $S_B$ on Alice's and Bob's side, respectively.
The protocol is correct when $S_A=S_B$. The resultant key is secret when $S_A$ is uniformly distributed and is uncorrelated with Eve's system. A protocol is called secure if it is both correct and secret. The protocol is robust if it never aborts when Eve is passive (i.e., Eve does not disturb the quantum channel) \cite{Finite-size-Furrer, Finite-size-Leverrier}.

However, for a real protocol we can only hope to achieve an almost ideal protocol up to small failure probabilities $\epsilon_{\rm cor}$ and $\epsilon_{\rm sec}$. The protocol is $\epsilon_{\rm cor}$-correct when $\Pr [{S_A} \ne {S_B}] \le {\epsilon _{\rm cor}}$. The protocol is $\epsilon_{\rm sec}$-secret when the key is $\delta$-close in trace distance to a uniformly distributed key that is uncorrelated with Eve's system, i.e., $\frac{1}{2}\left\| {{\rho _{{S_A}E'}} - {\tau _{{S_A}}} \otimes {\rho _{E'}}} \right\| \le {\delta} $, and $(1 - {p_{\rm abort}}){\delta} \le {\epsilon _{\rm sec}}$, where $\left\| . \right\|$ is the trace norm, and $p_{\rm abort}$ is the probability to abort. In this definition  $\tau _{S_A}$ is the uniform (i.e. fully mixed) state over $S_A$, $\rho _{E'}$ are states on Eve's system $E'$ (which characterizes Eve's quantum states $E$, as well as the public classical information $C$ leaked during the QKD protocol), and ${\rho _{{S_A}{E'}}} = \sum\limits_s {\left| s \right\rangle } \left\langle s \right| \otimes \rho _{E'}$ is a classical quantum state describing the state of $S_A$ and Eve's system $E'$ \cite{Finite-size-Furrer, Finite-size-Leverrier}.
A QKD protocol is called $\epsilon$-secure when it is $\epsilon_{\rm cor}$-correct and $\epsilon_{\rm sec}$-secret with $\epsilon_{\rm cor}+\epsilon_{\rm sec} \le \epsilon$ \cite{Finite-size-Furrer, Finite-size-Leverrier}. Note that this security definition also ensures that the QKD protocol is secure in the framework of composable security \cite{Finite-size-Furrer, Finite-size-Leverrier}, in which different cryptographic protocols can be combined without compromising the overall security.

Let us consider the equivalent entanglement-based scheme of a no-switching continuous-variable (CV) QKD protocol \cite{no-switching1,no-switching2} in the reverse reconciliation scheme where Alice prepares $N=2n$ two-mode squeezed vacuum states with the quadrature variance $V$, keeps the first mode of each state, while sending the second mode to Bob over an insecure quantum channel with transmissivity
$T$ and excess noise $\xi$ (relative to the input of the quantum channel). Alice and Bob measure their own modes with heterodyne detection to obtain two strings, $X \in {\mathbb{R}^{4n}}$ and $Y \in {\mathbb{R}^{4n}}$. Bob discretizes his string by dividing the continuous range of his quadrature variables $Y$ into $2^d$ intervals ${\mathcal{I}_1,...,\mathcal{I}_{2^d}}$ of the normal distribution, where $d$ is the discretization parameter. Bob applies the discretization map $\mathcal{D} : Y  \to U $, such that $\mathcal{D}(Y_k)=j$ if $Y_k \in \mathcal{I}_j$ \cite{Finite-size-Leverrier}. As a result of the discretization Bob ends up with the $m=4dn$-bit string $U$, where each symbol is encoded with $d$ bits of precision.

Here, similar to \cite{Finite-size-Leverrier} we assume parameter estimation can be performed after the reconciliation (or error correction). This assumption leads to the improvement of parameter estimation and enables us to use almost all the raw data to distill the secret key. In the error correction step based on a linear error-correcting code agreed on in advance Bob sends the syndrome of his vector $U$ of size $l_{\rm EC}$ to Alice, who outputs an estimate $\hat U$ of $U$. In order to know whether the error correction passed (i.e., $\hat U = U$), Alice and Bob compute a hash of their strings $\hat U$ and $U$, respectively. Bob then reveals his hash to Alice. If both hashes coincide, the protocol proceeds, otherwise it aborts. Note that the syndrome of size $l_{\rm EC}$ contributes to most of the leakage during the error correction. In the parameter estimation which is performed after the error correction, Bob sends only a few additional bits of information to Alice that allow her to compute the covariance matrix of the state $\rho_{AB}^ {\otimes(2n)}$ as well as a confidence region for the covariance matrix (for a detailed discussion on the parameter estimation and how Alice and Bob know the parameter estimation passed see \cite{Finite-size-Leverrier}). We indicate the maximum failure probabilities for the error correction and parameter estimation steps with $\epsilon_{\rm cor}$ and $\epsilon_{\rm PE}$. In the privacy amplification step Alice and Bob apply a random $\rm universal_2$ hash function to their respective strings, to extract two strings $S_A$ and $S_B$ of size $\ell$.

Based on the Leftover Hash Lemma \cite{lemma1,lemma2} the key of size $\ell$ is ${\epsilon_{\rm sec}}$-secret provided that $\ell$ is slightly smaller than the smooth min-entropy of Bob's string $U$ conditioned on Eve's system $E'$, $H_{\min}^{\epsilon_{\rm sm}}(U^m|E')$ \cite{lemma1}, where $m$ indicates the length of the string $U$, and ${\epsilon_{\rm sm}}$ is the smoothing parameter which is dependent on the value of ${\epsilon_{\rm sec}}$\footnote{In fact, the $\epsilon_{\rm sm}$-smooth min-entropy of Bob's string $U$ conditioned on Eve's system characterizes that given Eve's system how much $\epsilon_{\rm sm}$-close to uniform randomness (which is uncorrelated with Eve's system) can be extracted from the random variable $U$.}.
The conditional smooth min-entropy $H_{\min}^{\epsilon_{\rm sm}}(U^m|E')$ characterizes Eve's uncertainty (or lack of knowledge) about Bob's string $U$.
Note that the chain rule for the smooth min-entropy \cite{Finite-size-Leverrier} gives $H_{\min}^{\epsilon_{\rm sm}}(U^m|E')=H_{\min}^{\epsilon_{\rm sm}}(U^m|EC) \ge H_{\min}^{\epsilon_{\rm sm}}(U^m|E) - \log_2|C|$, where $\log_2|C|=l_{\rm EC}$.

In order to calculate the length $\ell$ of the final key which is $\epsilon$-secure, the conditional smooth min-entropy $H_{\min}^{\epsilon_{\rm sm}}(U^m|E)$ has to be lower bounded when the protocol did not abort, but this is usually a hard task. However, under the assumption of individual and collective attacks (meaning that every signal sent from Alice to Bob is attacked with the same quantum operation), where the state between Alice, Bob, and Eve has a tensor product structure, we can employ the Asymptotic Equipartition property \cite{Finite-size-Leverrier, Marco-thesis, Marco}, and provide a bound in terms of von Neumann entropy. This property states that for large $N$, the conditional smooth min-entropy approaches the conditional von Neumann entropy. Explicitly, we have $H_{\min }^{\epsilon_{\rm sm}} (U^m\left| E \right.) \ge H(U^m\left| E \right.) - \Delta_{\rm AEP}$ \cite{Finite-size-Leverrier}, where $H(U^m\left| E \right)$ is the conditional von Neumann entropy, and $\Delta_{\rm AEP} = \sqrt N [(d+1)^2+4(d+1)\sqrt{\log_2({2/\epsilon_{\rm sm}^2})} + 2\log_2({2}/({\epsilon^2 \epsilon_{\rm sm}})) ] + 4{\epsilon_{\rm sm}d}/{\epsilon}$ \cite{Finite-size-Leverrier,Finite-size-Lupo}. The conditional von Neumann entropy $H(U^m\left| E \right)$ is given by $H(U^m|E)=N H(U)- N \, \chi(U{:}E)$, where $H(U)$ is Bob's Shannon entropy, and  $\chi(U{:}E)$ is the Holevo mutual information between Eve and Bob for collective attacks. Note that for individual attacks $\chi(U{:}E)$ must be replaced by the Shannon mutual information between Eve and Bob, $I(U{:}E)$.


According to the security theorem proved in \cite{Finite-size-Leverrier,Finite-size-Lupo} the no-switching CV-QKD protocol is $\epsilon$-secure against collective attacks if \footnote{Note that terms here are slightly different to \cite{Finite-size-Leverrier} because, as pointed out in \cite{Finite-size-Lupo}, they are unnecessarily pessimistic on two counts. First, in Theorem 1 of Supplementary Information of \cite{Finite-size-Leverrier}, the terms $\epsilon_{\rm PE}$ and $\epsilon_{\rm cor}$ are both divided by $p$ (the unknown passing probability of the protocol) which is subsequently lower bounded by $\epsilon$, the overall collective security parameter. This is unnecessarily pessimistic and stems from substituting the unconditional failure probability for parameter estimation and error correction which are indeed  $\epsilon_{\rm PE}/p$ and $\epsilon_{\rm cor}/p$ respectively. However, the quantity in Eq.~(\ref{epsilon}) is conditioned upon passing the test, therefore the terms should be multiplied by $p$, which cancels. Secondly, in \cite{Finite-size-Leverrier} an extra step is introduced to bound the Shannon entropy appearing in Eq.~(\ref{key-length}) by the so-called empirical entropy. This leads to an extra correction term in Eq.~(\ref{key-length}) and an extra failure probability in Eq.~(\ref{epsilon}). However, neither of these are necessary since the term $N H(U) - l_{\rm EC}$ is directly measured in an experiment. Therefore it does not need to be rigorously bounded by the empirical entropy but can instead be modelled for the purposes of plotting the expected key rate by $N \beta I(A{:}B)$.}
\begin{equation}\label{epsilon}
\epsilon = 2\epsilon_{\rm sm} + \bar \epsilon + \epsilon_{\rm PE} + \epsilon_{\rm cor},
\end{equation}
and if the key length $\ell$ is chosen such that \footnote{Note that $\bar \epsilon$ comes from the Leftover Hash Lemma \cite{Finite-size-Leverrier}.}
\begin{equation}\label{key-length}
\begin{array}{l}
 \ell  \le N[ {{{ H}}(U) - \chi(U{:}E) } ] - {l _{\rm EC}}  - {\Delta _{\rm AEP}}  - 2\log_2 (\frac{1}{{2\bar \epsilon }}).
 \end{array}
\end{equation}
Considering that the leakage during the error correction is given by $l_{\rm EC} = N [H(U) - \beta I(A{:}B)]$ \cite{Finite-size-Leverrier,Finite-size-Lupo,Finite-size-Furrer}, where $I(A{:}B)$ is the Shannon mutual information between Alice and Bob, we can rewrite Eq.~(\ref{key-length}) as
\begin{equation}\label{key-length2}
\begin{array}{l}
 \ell  \le N[ \beta I(A{:}B) - \chi(U{:}E) ] - {\Delta _{\rm AEP}} - 2\log_2 (\frac{1}{{2\bar \epsilon }}),
 \end{array}
\end{equation}
where $\chi(U{:}E)$ is upper bounded by $\chi(Y{:}E)=\chi(B{:}E)$, since the discretization algorithm cannot increase the mutual information. According to \cite{Finite-size-Leverrier} the Holevo information $\chi(B{:}E)$ can be calculated based on a covariance matrix ${\bf{M}}_{ab} = [\sum _a^{\max }{\bf{I}},\sum _c^{\min }{\bf{Z}};\sum _c^{\min }{\bf{Z}},\sum _b^{\max }{\bf{I}}]$ with $\bf{I}$ a $2 \times 2$ identity matrix, and $\bf{Z} = \rm diag\left( {1, - 1} \right)$, where the elements of ${\bf{M}}_{ab}$ provide a bound on the elements of the covariance matrix of the state shared between Alice and Bob:
\begin{equation}\label{bound-CM}
\begin{array}{l}
\sum _a^{\max } = \frac{1}{{2n}}\left[ {1 + 2\sqrt {\frac{{\log (36/{\epsilon _{\rm PE}})}}{n}} } \right]{\left\| X \right\|^2} - 1,\\
\\
\sum _b^{\max } = \frac{1}{{2n}}\left[ {1 + 2\sqrt {\frac{{\log (36/{\epsilon _{\rm PE}})}}{n}} } \right]{\left\| Y \right\|^2} - 1,\\
\\
\sum _c^{\min } = \frac{1}{{2n}}\left\langle {X,Y} \right\rangle  - 5\sqrt {\frac{{\log (8/{\epsilon _{\rm PE}})}}{{{n^3}}}} \left( {{{\left\| X \right\|}^2} + {{\left\| Y \right\|}^2}} \right),
\end{array}
\end{equation}
where ${\left\| X \right\|^2}, {\left\| Y \right\|^2}, \left\langle {X,Y} \right\rangle$ can be achieved by taking values differing by 3 standard deviations from the expected values \cite{Finite-size-Leverrier} (for an expected Gaussian channel with parameters $T$ and $\xi$). It is then assumed Eve's information can be upper bounded by calculating $\chi(B{:}E)$ based on the covariance matrix ${\bf{M}}_{ab}$, except with the probability of $\epsilon_{\rm PE}$.


The final key rate where the key is $\epsilon$-secure against collective attacks is given by $\ell/N$. We recall that in Eq.~(\ref{key-length2}) we have considered the same scenario as \cite{Finite-size-Leverrier}, where almost all the raw data can be utilized to distill the secret key\footnote{Note that it has been recently shown in \cite{PE-MDI-2018} that in CV-QKD the whole raw keys can be used for both parameter estimation and secret key generation, without compromising the security, and without any requirements of doing error correction before parameter estimation.} (by performing the parameter estimation after the error correction). However, if the parameter estimation is performed before the error correction, Alice and Bob are required to disclose non-negligible bits of information, $N_{\rm PE}$, during the parameter estimation, which means $N'$ bits of information is used for the key extraction, where $N'=N-N_{\rm PE}$. As a result, the final secure key rate is given by $\ell/N$, where $\ell$ is given by Eq.~(\ref{key-length2}), but now $N$ in Eq.~(\ref{key-length2}) has to be replaced by $N'$.

In order to prove the security of the no-switching CV-QKD protocol against coherent attacks we apply the Gaussian de Finetti reduction technique \cite{Finite-size-Leverrier2017}. In order to apply this technique we need to truncate the Hilbert space in a suitable manner. This can be achieved with the help of an energy test \cite{Finite-size-Leverrier2017}, which ensures that the state shared between Alice and Bob is suitably described by assigning a low-dimensional Hilbert space. Considering the input state is a $(2n+k)$-mode state, Alice and Bob should symmetrize this state and measure the last $k$ modes with heterodyne detection. If the average energy per mode is below $d_A$ for Alice and $d_B$ for Bob, the energy test passes and Alice and Bob apply the CV-QKD to their remaining modes, otherwise the protocol aborts. The thresholds $d_A$ and $d_B$ should be chosen properly to ensure that the energy test passes with large success probability.

According to the security theorem proved in \cite{Finite-size-Leverrier2017}, if we are given a no-switching CV-QKD protocol acting on $2n$-mode state shared between Alice and Bob (which is suitably symmetrized), such that the protocol is $\epsilon$-secure against Gaussian collective attacks, the modified protocol including an energy test and an additional privacy amplification step \cite{Finite-size-Leverrier2017} is $\tilde \epsilon$-secure against coherent attacks, with $\tilde \epsilon = (K^4/50) \epsilon$ where
\begin{equation}\label{epsilon-tilde}
\begin{array}{l}
 K = \max \left[ {1,n({d_A} + {d_B})\left( {1 + 2\sqrt {[(\ln (8/\epsilon ))/2n]} } \right.} \right. \\
  \\
 \left. { + \,\left. {(\ln (8/\epsilon ))/n} \right){{\left( {1 - 2\sqrt {[(\ln (8/\epsilon ))/2k]} } \right)}^{ - 1}}} \right]. \\
 \end{array}
\end{equation}
Thus, the security loss due to the reduction from coherent attacks to collective attacks scales like $\mathit{O}(n^4)$.

Note that in our numerical calculations of finite-size key rate we consider the security parameter of $\tilde \epsilon = 10 ^ {-6}$ for all types of attacks. Thus, we choose $\epsilon_{\rm PE}=\epsilon_{\rm cor}=\bar \epsilon=10^{-7}$ for individual attacks to satisfy Eq.~(\ref{epsilon}) for $\epsilon=\tilde \epsilon$. For coherent attacks (which are modelled by Gaussian collective attacks with far smaller security parameter $\epsilon=10^{-42} \ll \tilde \epsilon = 10 ^ {-6}$ for $n=10^9$) we choose $\epsilon_{\rm PE}=\epsilon_{\rm cor}=\bar \epsilon=10^{-43}$ to satisfy Eq.~(\ref{epsilon}). However, for hybrid attacks we consider a pessimistic scenario by choosing $\epsilon_{\rm PE}=\epsilon_{\rm cor}=\bar \epsilon=10^{-43}$, which again leads to a loose lower bound on the finite-size hybrid key rate.

Note also that in our numerical calculations of finite-size key rate we do not use directly the covariance matrix shared between Alice and Bob given by ${\bf{M}}_{ab}$ to compute the key rate. More specifically, we first calculate the matrix ${\bf{M}}_{ab}$, and then estimate the required parameters ($T$, $\xi$, and $V$) from the elements of the matrix ${\bf{M}}_{ab}$, and then proceed to compute the key rate based on the calculations provided in the next sections.

\section{Appendix B: Calculation of $\boldsymbol{I(A:B)}$}\label{APP.B}

In the entanglement-based scheme of the no-switching CV-QKD protocol the initial pure Gaussian entangled state $\rho_{AB}$ with the quadrature variance $V$ is completely described by its first moment, which is zero, and its covariance matrix,
\begin{eqnarray}\label{AB-CM}
{\bf{M}}_{AB}= \left[ {\begin{array}{*{20}{c}}
{V\,\bf{I}}&{\sqrt {{V^2} - 1} \,\bf{Z}}\\
{\sqrt {{V^2} - 1} \,\bf{Z}}&{V\,\bf{I}}
\end{array}} \right] .
\end{eqnarray}
 After transmission of mode~$B$ through a quantum channel with transmissivity $T$ and excess noise $\xi$, the covariance matrix of the mixed state ${\rho _{A{B_1}}}$ at the output of the channel is given by
\begin{equation}\label{AB1-CM}
{{\bf{M}}_{A{B_1}}} = \left[ {\begin{array}{*{20}{c}}
{V\,\bf{I}}&{\sqrt {T \,} \sqrt {{V^2} - 1}\,\bf{Z}}\\
{\sqrt T  \,\sqrt {{V^2} - 1}\,\bf{Z}}&{\left( {T (V + {\chi _{\rm line}})} \right)\,\bf{I}}
\end{array}} \right],
\end{equation}
where ${\chi _{\rm line}} = \xi  + \frac{{1}}{T}-1$. At the output of the channel Bob applies heterodyne detection to mode~$B_1$. Bob's heterodyne detector with efficiency of $\eta$ and electronic noise variance of $\upsilon_{\rm el}$ can be modeled by placing a beam splitter of transmissivity $\eta$ before an ideal heterodyne detector \cite{inefficient_homodyne, inefficient_heterodyne}. The heterodyne detector's electronic noise can be modelled by a two-mode squeezed vacuum state, $\rho_{{F_0}G}$, of quadrature variance $\upsilon $, where $\upsilon  = 1 + {2\upsilon _{\rm el}}/(1 - \eta )$. One input port of the beam splitter is the received mode $B_1$, and the second input port is fed by one half of the entangled state $\rho_{{F_0}G}$, mode~$F_0$, while the output ports are mode~$B_{2}$ (which is measured by the ideal heterodyne detector) and mode $F$. 

The Shannon mutual information between Alice and Bob, $I(A:B)$, is given by $I(A:B) = {\log _2}\frac{{{V_{B_2^{\rm het}}}}}{{{V_{{B_2}^{\rm het}\left| A^{\rm het} \right.}}}}$, where $V_{B_2^{\rm het}}$ is the variance of heterodyne-detected mode $B_2$, and is given by $V_{B_2^{\rm het}}=\eta T (V+\chi_{\rm tot})/2$, where $\chi_{\rm tot}  = {\chi _{\rm line}} + \frac{{{\chi _{\rm het}}}}{T }$, and where $\chi_{\rm het} = [1+(1 - \eta ) + {2\upsilon_{\rm el}}]/\eta $. The conditional variance $V_{{B_2}^{\rm het}|A^{\rm het}}$ is the variance of heterodyne-detected mode $B_2$ conditioned on Alice's heterodyne detection of mode $A$, which is given by $V_{{B_2}^{\rm het}|A^{\rm het}} = \eta T (1+\chi_{\rm tot})/2$.

\section{Appendix C: Calculation of $\boldsymbol{I_\mu^{\rm hyb}(B:E)}$}\label{APP.C}

The upper bound on the mutual information between Eve and Bob in the hybrid attack, $I_\mu^{\rm hyb}(B:E)$, is given by 
\begin{equation}\label{mutual-hybrid-appendix}
\begin{array}{l}
I_\mu^{\rm hyb}(B:E) = I_\mu^{\rm hyb}(B:E'_1E'_2 E''_1E''_2 ) =
\\
\\
\chi_\mu(BE''_1E''_2:E'_1E'_2) + I_\mu(B:E''_1E''_2) - \chi_\mu(E''_1E''_2:E'_1E'_2).
 \end{array}
\end{equation}
We now analyse the calculation of the mutual information terms on the right-hand-side of Eq.~(\ref{mutual-hybrid-appendix}):
\subsection{1. Calculation of $\boldsymbol{\chi_\mu(BE''_1E''_2:E'_1E'_2)}$}

In Eq.~(\ref{mutual-hybrid-appendix}) the Holevo mutual information $\chi_\mu(BE''_1E''_2:E'_1E'_2)$ is given by $\chi_\mu(BE''_1E''_2:E'_1E'_2) = S(\rho_{E'_1E'_2})-S(\rho_{E'_2E'_2|E''_1E''_2B_2})$, where $S(\rho)$ is the von Neumann entropy\footnote{The von Neumann entropy of an $n$-mode Gaussian state $\rho$ with the covariance matrix $\bf{M}$ is given by $S(\rho)=\sum\nolimits_{i = 1}^n {G(\frac{\lambda _i-1}{2})} $, where $\lambda _{i}$ are the symplectic eigenvalues of the covariance matrix ${\bf{M}}$, and $G(x) = (x + 1){\log _2}(x + 1) - x {\log _2}(x)$.} of the state $\rho$. Note that here we assume Bob's detection noise is not accessible to Eve. The first entropy $S(\rho_{E'_1E'_2})$ is calculated through the symplectic eigenvalues of the covariance matrix ${\bf{M}}_{E'_1E'_2}$, which is given by
\begin{equation}\label{Eve-col-CM}
\begin{array}{l}
 {\bf{M}}_{{{E_1'}}{{E_2'}}} =  \\
 \\
 \left[ {\begin{array}{*{20}{c}}
   {\left[ {{\tau _1}{V_{{E^c_1}}} + (1 - {\tau _1}){\omega _1}} \right]\bf{I}} & {\sqrt {{\tau _1}{\tau _2}} {C_{{E^c_1},{E^c_2}}}\bf{Z}}  \\
   {\sqrt {{\tau _1}{\tau _2}} {C_{{E^c_1},{E^c_2}}}\bf{Z}} & {\left[ {{\tau _2}{V_{{E^c_2}}} + (1 - {\tau _2}){\omega _2}} \right]\bf{I}}  \\
\end{array}} \right], \\
 \end{array}
\end{equation}
where $V_{E^c_1} = \mu V_{E_1} + (1-\mu)$, $V_{E^c_2} = \mu V_{E_2} + (1-\mu)$, and $C_{E^c_1,E^c_2} = \mu C_{E_1,E_2}$. Note that $V_{E_1} = T\omega_E + (1-T)V$, $V_{E_2} = \omega_E $, and $C_{E_1,E_2} = \sqrt{T}\sqrt{\omega_E^2-1}$.\hfill\break
The second entropy we require in order to determine $\chi_\mu(BE''_1E''_2:E'_1E'_2)$ is $S(\rho_{E'_2E'_2|E''_1E''_2B_2})$ which is calculated through the symplectic eigenvalues of the conditional covariance matrix ${\bf{M}}_{E'_1E'_2|E''_1E''_2B_2}$. This conditional covariance matrix is actually the covariance matrix of modes~$E'_1E'_2$ conditioned on the homodyne detection of modes~$E''_1E''_2$ and heterodyne detection of mode~$B_2$. \hfill\break
Let us recall that the heterodyne detection of mode~$B_2$ is the combination of mode~$B_2$ with a vacuum mode in a balanced beam splitter, which outputs mode~$B_3$ and mode~$C$, where the $\hat q$ ($\hat p$) quadrature is measured on mode~$B_3$ (mode~$C$) using a homodyne detector. Hence, the heterodyne detection on mode~$B_2$ is actually a conjugate homodyne detection on modes~$B_3$ and $C$. In this case we have ${\bf{M}}_{E'_1E'_2|E''_1E''_2B_2}={\bf{M}}_{E'_1E'_2|E''_1E''_2B_3C}$, where we have
\begin{equation}\label{CM-E'_1E'_2|E''_1E''_2B_3C}
\begin{array}{l}
{\bf{M}}_{E'_1E'_2|E''_1E''_2B_3C}=\\
{\bf{M}}_{E'_1E'_2}-\boldsymbol{\sigma}_{E'_1E'_2,E''_1E''_2{B_3}C} \,\, {\bf{H}}_{\rm hom} \,\, \boldsymbol{\sigma}^T_{E'_1E'_2,E''_1E''_2{B_3}C}.
 \end{array}
\end{equation}
In Eq.~(\ref{CM-E'_1E'_2|E''_1E''_2B_3C}) the covariance matrix ${\bf{M}}_{E'_1E'_2}$ is given by Eq.~(\ref{Eve-col-CM}), and the matrix $\boldsymbol{\sigma}_{E'_1E'_2,E''_1E''_2{B_3}C}$ is given by
\begin{equation}\label{sigma-E'_1E'_2,E''_1E''_2B_3C}
\boldsymbol{\sigma}_{E'_1E'_2,E''_1E''_2{B_3}C} = \left[ {\begin{array}{*{20}{c}}
   {\boldsymbol{\sigma}_{E'_1E'_2,E''_1E''_2}} & {\boldsymbol{\sigma}_{E'_1E'_2,{B_3}}} & {\boldsymbol{\sigma}_{E'_1E'_2,C}} \\
   \end{array}} \right].
\end{equation}
In Eq.~(\ref{sigma-E'_1E'_2,E''_1E''_2B_3C}) the matrix $\boldsymbol{\sigma}_{E'_1E'_2,E''_1E''_2}$ is given by
\begin{equation}\label{sigma-E'_1E'_2,E''_1E''_2}
\begin{array}{l}
\boldsymbol{\sigma}_{E'_1E'_2,E''_1E''_2} = \\
\\
\left[ {\begin{array}{*{20}{c}}
{{C_{{q_{{{E_1'}}}},{q_{{{E_1''}}}}}}}&0&{{C_{{q_{{{E_1'}}}},{q_{{{E_2''}}}}}}}&0\\
0&{{C_{{p_{{{E_1'}}}},{p_{{{E_1''}}}}}}}&0&{{C_{{p_{{{E_1'}}}},{p_{{{E_2''}}}}}}}\\
{{C_{{q_{{{E_2'}}}},{q_{{{E_1''}}}}}}}&0&{{C_{{q_{{{E_2'}}}},{q_{{{E_2''}}}}}}}&0\\
0&{{C_{{p_{{{E_2'}}}},{p_{{{E_1''}}}}}}}&0&{{C_{{p_{{{E_2'}}}},{p_{{{E_2''}}}}}}}
\end{array}} \right].
 \end{array}
\end{equation}
In Eq.~(\ref{sigma-E'_1E'_2,E''_1E''_2}) we have $C_{q_{E_1'},q_{E_1''}} = C_{p_{E_1'},p_{E_2''}} = (C_{E^i_1,E_1'} - C_{E^i_2,E_1'}) / \sqrt{2}$, $C_{p_{E_1'},p_{E_1''}} = C_{q_{E_1'},q_{E_2''}} = (C_{E^i_1,E_1'} + C_{E^i_2,E_1'}) / \sqrt{2}$, $C_{q_{E_2'},q_{E_2''}} = - C_{p_{E_2'},p_{E_1''}} = (C_{E^i_1,E_2'} + C_{E^i_2,E_2'}) / \sqrt{2}$, and $C_{p_{E_2'},p_{E_2''}} = - C_{q_{E_2'},q_{E_1''}} = (- C_{E^i_1,E_2'} + C_{E^i_2,E_2'}) / \sqrt{2}$,
and where
\begin{equation}\label{var-cov1}
\begin{array}{*{20}{c}}
C_{E^i_1,E_1'} = \sqrt{\tau_1(1-\mu)\mu}(1 - V_{E_1}),
\\
\\
C_{E^i_2,E_1'} = -\sqrt{\tau_1(1-\mu)\mu}C_{E_1,E_2},
\\
\\
C_{E^i_1,E_2'} = -\sqrt{\tau_2(1-\mu)\mu}C_{E_1,E_2},
\\
\\
C_{E^i_2,E_2'} = \sqrt{\tau_2(1-\mu)\mu}(1 - V_{E_2}).
\end{array}
\end{equation}
In Eq.~(\ref{sigma-E'_1E'_2,E''_1E''_2B_3C}) the matrices $\boldsymbol{\sigma}_{E'_1E'_2,{B_3}}$ and $\boldsymbol{\sigma}_{E'_1E'_2,C}$ are given by $\boldsymbol{\sigma}_{E'_1E'_2,{B_3}} = -\boldsymbol{\sigma}_{E'_1E'_2,C} = \frac{1}{\sqrt{2}}\boldsymbol{\sigma}_{E'_1E'_2,{B_2}}$, where we have
\begin{equation}\label{sigma-E'_1E'_2,B_2}
\boldsymbol{\sigma}_{E'_1E'_2,{B_2}} = \left[ {\begin{array}{*{20}{c}}
   {\sqrt {{\tau _1} \mu (1 - T)T\eta} \left( {{\omega _E} - V} \right) \bf{I}}  \\
   {\sqrt {{\tau _2} \mu (1 - T)\eta} \sqrt {\omega _E^2 - 1} \bf{Z}}  \\
\end{array}} \right].
\end{equation}
In Eq.~(\ref{CM-E'_1E'_2|E''_1E''_2B_3C})  the matrix ${\bf{H}}_{\rm hom}$ is given by ${\bf{H}}_{\rm hom} = ({\bf{X}} {\bf{M}}_{E''_1E''_2{B_3}C} {\bf{X}})^{\rm MP}$, where ${\bf{X}}=\rm diag(1,0,0,1,1,0,0,1)$, MP stands for the Moore-Penrose pseudoinverse of a matrix, and the covariance matrix ${\bf{M}}_{E''_1E''_2{B_3}C}$ is given by
\begin{equation}\label{CM-E''_1E''_2B_3C}
{\bf{M}}_{E''_1E''_2{B_3}C} = \left[ {\begin{array}{*{20}{c}}
   {{\bf{M}}_{E''_1E''_2}} & {\boldsymbol{\sigma}^T_{{B_3},E''_1E''_2}} & {\boldsymbol{\sigma}^T_{C,E''_1E''_2}} \\
   {\boldsymbol{\sigma}_{{B_3},E''_1E''_2}} & {{\bf{M}}_{B_3}} & {\boldsymbol{\sigma}^T_{C,{B_3}}}\\
   {\boldsymbol{\sigma}_{C,E''_1E''_2}} & {\boldsymbol{\sigma}_{C,{B_3}}} & {{\bf{M}}_C}\\
   \end{array}} \right].
\end{equation}
In Eq.~(\ref{CM-E''_1E''_2B_3C}) the covariance matrix ${\bf{M}}_{E''_1E''_2}$ is given by
\begin{equation}\label{Eve-ind-CM}
\begin{array}{l}
{\bf{M}}_{E''_1E''_2} = \\
\\
\left[ {\begin{array}{*{20}{c}}
{{V_{{q_{{{E_1''}}}}}}}&0&{{C_{{q_{{{E_1''}}}},{q_{{{E_2''}}}}}}}&0\\
0&{{V_{{p_{{{E_1''}}}}}}}&0&{{C_{{p_{{{E_1''}}}},{p_{{{E_2''}}}}}}}\\
{{C_{{q_{{{E_1''}}}},{q_{{{E_2''}}}}}}}&0&{{V_{{q_{{{E_2''}}}}}}}&0\\
0&{{C_{{p_{{{E_1''}}}},{p_{{{E_2''}}}}}}}&0&{{V_{{p_{{{E_2''}}}}}}}
\end{array}} \right],
\end{array}
\end{equation}
where $V_{q_{E''_1}} = V_{p_{E''_2}}=(V_{E^i_1}+V_{E^i_2})/2-C_{{E^i_1},{E^i_2}}$, $V_{p_{E''_1}}=V_{q_{E''_2}}=(V_{E^i_1}+V_{E^i_2})/2+C_{{E^i_1},{E^i_2}}$, and $C_{q_{E''_1},q_{E''_2}}=C_{p_{E''_1},p_{E''_2}}=(V_{E^i_1}-V_{E^i_2})/2$, and where
\begin{equation}\label{var-Eve-ind-CM}
\begin{array}{l}
V_{E^i_1} = (1-\mu) V_{E_1} + \mu ,\,\,\,\ V_{E^i_2} = (1-\mu) V_{E_2} + \mu,
\\
\\
C_{{E^i_1},{E^i_2}} = (1-\mu) C_{E_1,E_2} .
\end{array}
\end{equation}
In Eq.~(\ref{CM-E''_1E''_2B_3C}) the matrices $\boldsymbol{\sigma}_{{B_3},E''_1E''_2}$ and $\boldsymbol{\sigma}_{C,E''_1E''_2}$ are given by $\boldsymbol{\sigma}_{{B_3},E''_1E''_2} = -\boldsymbol{\sigma}_{C,E''_1E''_2} = \frac{1}{\sqrt{2}}\boldsymbol{\sigma}_{{B_2},E''_1E''_2}$, where
\begin{equation}\label{sigma-B_2,E''_1E''_2}
\begin{array}{l}
\boldsymbol{\sigma}_{B_2,E''_1E''_2} = \\
\\
\left[ {\begin{array}{*{20}{c}}
{{C_{{q_{{B_2}}},{q_{{{E_1''}}}}}}}&0&{{C_{{q_{{B_2}}},{q_{{{E_2''}}}}}}}&0\\
0&{{C_{{p_{{B_2}}},{p_{{{E_1''}}}}}}}&0&{{C_{{p_{{B_2}}},{p_{{{E_2''}}}}}}}
\end{array}} \right].
\end{array}
\end{equation}
In Eq.~(\ref{sigma-B_2,E''_1E''_2}) we have $C_{q_{B_2},q_{E''_1}}=C_{p_{B_2},p_{E''_2}}=(C_{{B_2},{E^i_1}} - C_{{B_2},{E^i_2}})/\sqrt{2} $, and $C_{p_{B_2},p_{E''_1}}=C_{q_{B_2},q_{E''_2}} = (C_{{B_2},{E^i_1}} + C_{{B_2},{E^i_2}})/\sqrt{2}$, and where
\begin{equation}\label{var-cov2}
\begin{array}{*{20}{c}}
C_{{B_2},{E^i_1}} = \sqrt{(1-\mu)(1-T)T\eta}  (V - \omega_E),
\\
\\
C_{{B_2},{E^i_2}} = - \sqrt{(1-\mu)(1-T)\eta}  \sqrt{\omega_E^2-1} .
\end{array}
\end{equation}
In Eq.~(\ref{CM-E''_1E''_2B_3C}) we have ${\bf{M}}_{B_3} = {\bf{M}}_{C} = 0.5 (V_{B_2}+1) \bf{I}$, and $\boldsymbol{\sigma}_{C,{B_3}}=0.5 (1-V_{B_2}) \bf{I}$, where $V_{B_2} = {\eta T (V + \chi_{\rm t} )}$ and where $\chi_{\rm t}  = {\chi _{\rm line}} + \frac{{{\chi _{D}}}}{T }$, and where $\chi_{D} = [(1 - \eta ) + {2\upsilon_{\rm el}}]/\eta $.

\subsection{2. Calculation of $\boldsymbol{I_\mu(B:E''_1E''_2)}$}

In Eq.~(\ref{mutual-hybrid-appendix}) the Shannon mutual information $I_\mu(B:E''_1E''_2)$ is given by
\begin{equation}\label{mutual-Bob-Ind}
I_\mu(B:E''_1E''_2) = \frac{1}{2} \log_2\frac{V_{B_2^{\rm het}}}{V_{q_{B^{\rm het}_2}|q_{E''_1}}} + \frac{1}{2} \log_2\frac{V_{B_2^{\rm het}}}{V_{p_{B^{\rm het}_2}|p_{E''_2}}},
\end{equation}
where the conditional variance $V_{q_{B^{\rm het}_2}|q_{E''_1}}$ is given by $V_{q_{B^{\rm het}_2}|q_{E''_1}}{=}(V_{q_{B_2}|q_{E''_1}}{+}1)/2$, and similarly for the $p$ quadrature we have $V_{p_{B^{\rm het}_2}|p_{E''_2}}{=}(V_{p_{B_2}|p_{E''_2}}{+}1)/2$. The symmetry of Eve's information on $q_{B_2}$ and $p_{B_2}$ imposes that $V_{q_{B_2}|q_{E''_1}}{=}V_{p_{B_2}|p_{E''_2}} $. Note that  $V_{q_{B_2}|q_{E''_1}} {=} V_{q_{B_2}} {-} C_{q_{B_2},q_{E''_1}}/V_{q_{E''_1}}$, where $V_{q_{B_2}}{=}\eta T (V + \chi_{\rm t} )$. 
Note also that $V_{B_2^{\rm het}}$, $C_{q_{B_2},q_{E''_1}}$, and $V_{q_{E''_1}}$ have been already calculated and provided in the previous sections.

\subsection{3. Calculation of $\boldsymbol{\chi_\mu(E''_1E''_2:E'_1E'_2)}$}

In Eq.~(\ref{mutual-hybrid-appendix}) the Holevo information $\chi_\mu(E''_1E''_2:E'_1E'_2)$ is given by $\chi_\mu(E''_1E''_2:E'_1E'_2) = S(\rho_{E'_1E'_2})-S(\rho_{E'_1E'_2|E''_1E''_2})$. The conditional entropy $S(\rho_{E'_1E'_2|E''_1E''_2})$ is calculated through the symplectic eigenvalues of the conditional covariance matrix ${\bf{M}}_{E'_1E'_2|E''_1E''_2}$. This conditional covariance matrix is given by ${\bf{M}}_{E'_1E'_2|E''_1E''_2}={\bf{M}}_{E'_1E'_2}-\boldsymbol{\sigma}_{E'_1E'_2,{E''_1E''_2}} \,\, {\bf{H}}^i_{\rm hom} \,\, \boldsymbol{\sigma}^T_{E'_1E'_2,{E''_1E''_2}}$. The matrix ${\bf{H}}^i_{\rm hom}$ is given by ${\bf{H}}^i_{\rm hom} = ({\bf{X}}_i {\bf{M}}_{E''_1E''_2} {\bf{X}}_i)^{\rm MP}$, where ${\bf{X}}_i=\rm diag(1,0,0,1)$. Note that the matrices ${\bf{M}}_{E'_1E'_2}$, $\boldsymbol{\sigma}_{E'_1E'_2,{E''_1E''_2}}$, and ${\bf{M}}_{E''_1E''_2}$ are given by Eqs.~(\ref{Eve-col-CM}), (\ref{sigma-E'_1E'_2,E''_1E''_2}), and (\ref{Eve-ind-CM}), respectively.

\begin{figure}[t!]
    \begin{center}
      {\includegraphics[width=3.5 in]{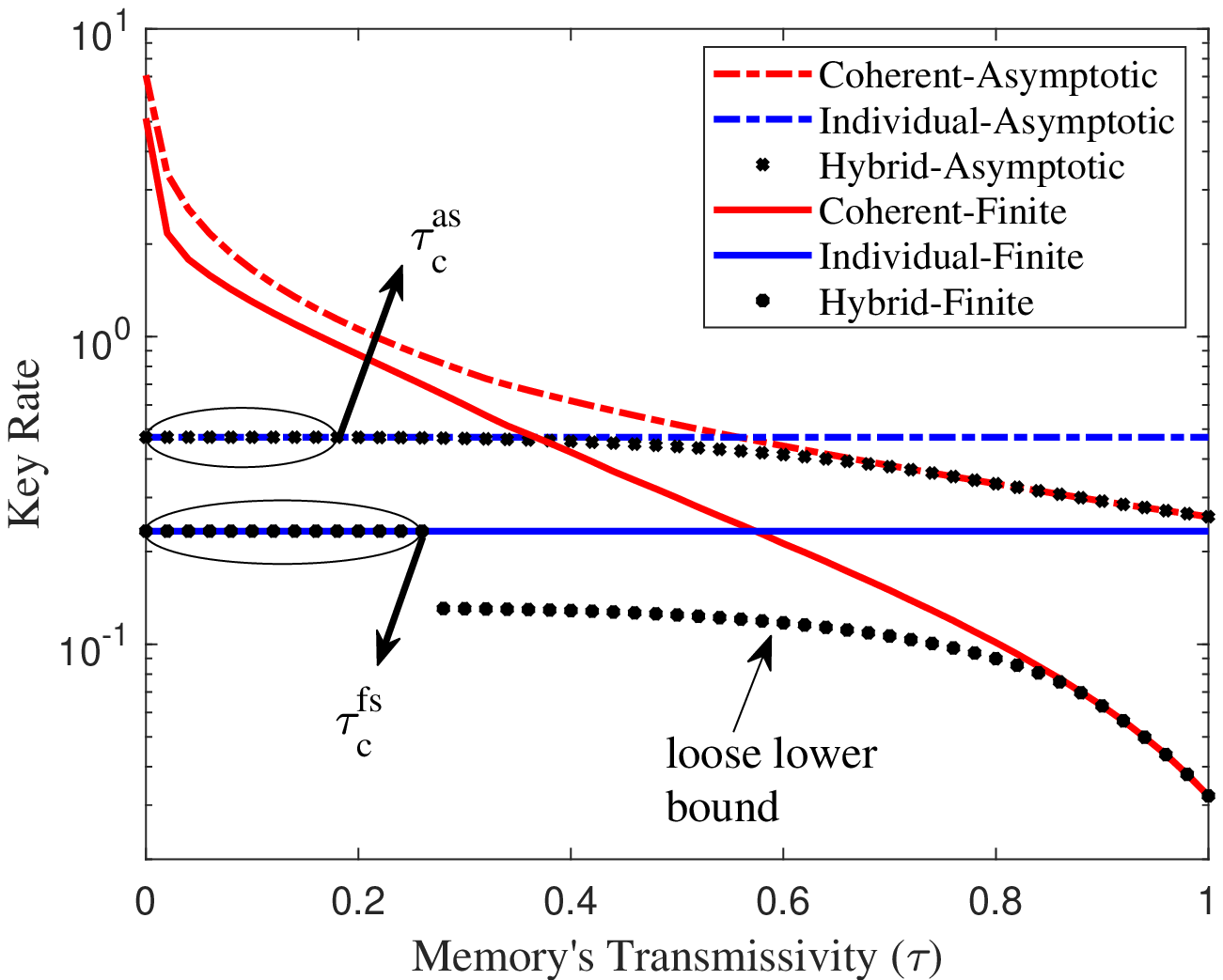}}
    \caption{The finite-size and asymptotic key rate as a function of memory's transmissivity, $\tau$, for individual (blue), coherent (red), and hybrid attacks (black). The numerical values are $\eta{=}0.6$ \cite{experiment-CVQKD-2016}, $\upsilon_{\rm el}{=}0.015$ \cite{experiment-CVQKD-2013}, $T{=}0.5$, $\xi{=}0.01$, $d{=}5$, $\omega{=}1, \beta{=}0.98$ \cite{beta-0.98},  and the modulation variance is optimized. The region marked by the ellipse shows memory's transmissivities for which Eve's optimal attack is the individual attack.}\label{FS}
    \end{center}
\end{figure}

\subsection{Appendix D: Numerical results on low-loss channels}\label{APP.D}

We repeat our numerical simulations for a low-loss quantum channel in Fig.~\ref{FS}, which result in similar trends to high-loss channels.  Asymptotically we see that when $\tau{\le}\tau_c^{\rm as} {=} 0.18$ individual attacks are optimal, for $\tau_c^{\rm as}{<}\tau{\le}0.75$ Eve's optimal strategy is a hybrid combination of both individual and coherent attacks, and for $\tau{>}0.75$ coherent attacks are optimal. In the finite-size case when $\tau{\le}\tau^{\rm fs}_{c}{=}0.26$ individual attacks are optimal, for $\tau_c^{\rm fs}{<}\tau{\le}0.89$ Eve's optimal strategy is a hybrid combination of both individual and coherent attacks (only a loose lower bound on the finite-size hybrid key rate can be calculated), and for $\tau{>}0.89$ coherent attacks are optimal. Note that in Fig.~\ref{FS} the threshold transmissivity is different for the finite-size and asymptotic regime, since the finite key rate is calculated based on the estimated values of the channel, while the asymptotic key rate is calculated based on the expected values of the channel.

Additional calculations beyond those illustrated here have been carried out covering direct reconciliation (DR), which result in similar trends to those indicated here. However, DR is only successful when the channel loss is below 3dB. For instance, in the DR scenario of the no-switching protocol with the same parameters as Fig.~\ref{FS}, individual attacks result in positive finite key rates only for low-loss channels with $T \ge 0.72$. Hence, if Eve's decoherence is large enough to make individual attacks the optimal attacks, positive finite key rates can only be generated for $T \ge 0.72$.



\end{document}